\documentclass[twocolumn,showpacs,preprintnumbers,amsmath,amssymb]{revtex4}

\usepackage{graphicx}
\usepackage{dcolumn}
\usepackage{bm}

\begin{document}

\preprint{}

\title{Quantum Monte Carlo Simulation of two-dimensional Emery model}

\author{Bernard Martinie}
\affiliation{
D\'epartement de Physique, UFR Sciences et Techniques, Parc Grandmont, 37200 Tours, France\\
}

\date{\today}

\begin{abstract}
The Quantum Monte Carlo simulation of the two-dimensional Emery model of CuO$_{2}$  plane of hight T$_{c}$  oxide superconductors were performed. The method based on the direct-space proposed by Suzuki and al., Hirsch and al. was used. Contrary to the method based on the Hubbard-Stratonovich transformation, the states generated by this method are basis states in occupation number representation, i. e. configurations of fermions can be observed on real two-dimensional array. Energy and specific heat were computed for different dopings during decreasing temperature. The specific heat curves show peaks at low temperature which could be assigned to electronic transitions. Quantity similar to current-current correlation function were computed. The static electric conductivity curves obtained by this way show metal-insulator transitions for doping $\delta=0$ and doping $\delta=1$, and two different metallic behaviours for intermediate dopings. On the direct-space states generated at low temperature and doping $\delta=0$, the fermions form antiferromagnetic loops while they form antiferromagnetic chains for other dopings. The loops can be related to circulating currents and the chains to the stripes. The antiferromagnetic loops seem to appear when the conductivity becomes zero while the conductivity increase with the numbers of chains but superconductivity is not unambiguously evident. 
\end{abstract}

\maketitle

\section{Introduction}
The Emery model\cite{Emery1} of the CuO$_{2}$ plane was proposed and extensively studied to explain the behaviour of hight T$_{c}$ oxide superconductors. This model seems to reproduce the essential phenonema observed on materials although the evidence of superconductivity is not totally proven for the repulsive Emery model. Many results are obtained by numerical simulations using different quantum Monte Carlo methods. These methods are based on the Hubbard-Stratonovich  transformation\cite{Blanckenbecler1}\cite{White1} which does not generate actual fermions configurations.
\\ In this paper, we present results on two-dimensional repulsive Emery model obtained with a method based on the direct-space proposed by Suzuki and al.\cite{Suzuki1}\cite{Suzuki2} and Hirsch and al.\cite{Hirsch1}\cite{Hirsch2}. At fixed temperature, this method allows to generate  some of the most representative occupation number basis states of the model. These states are used to compute average values of energy, specific heat and rough static electric conductivity.
\\The paper is organized as follows:
\begin{itemize}
\item in section 2 the Emery model is recalled,
\item in section 3 the numerical method is described,
\item in section 4 the validity of the method is tested with the two-dimensional Hubbard model,
\item in section 5 the energy and specific heat curves are shown, and some direct-space states are presented,
\item in section 6 we define a way for computing a rought value of conductivity and present the computed curves,
\item in section 7 we analyze the results,
\item in section 8 some concluding remarks are given.
\end{itemize}

\section{Emery model}
\begin{figure}
\begin{center}
\includegraphics[width=6cm]{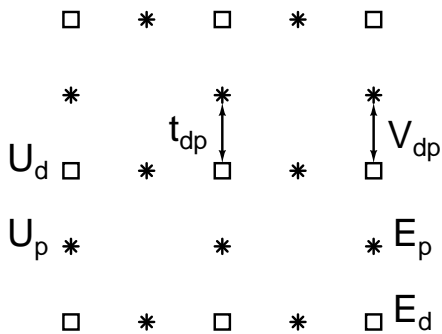}
\end{center}
\caption{The structure of the CuO$_{2}$ plane. The squares indicate Cu sites and the stars oxygen sites.}
\label{fig:Emerymodel}
\end{figure}

The studied system is the CuO$_{2}$ plane which have the structure shown in Fig.  \ref{fig:Emerymodel}. One considers the behaviour of holes on this array and uses the occupation number representation. The single-particle states used to construct the basis states of the representation are the hole Wannier states localised on each site. According to atomic orbitals involved for copper and oxygen atoms, the Wannier state on all copper sites are labeled \textbf{d} while Wannier states on all oxygen sites are labeled \textbf{p}. It will be assumed that the Hamiltonian of the system is an extended Hubbard Hamiltonian
\begin{eqnarray}
\label{hamiltonian1} 
H =&&t_{dp}\sum_{\left\langle i,j\right\rangle,\sigma }\left( d^{\dagger}_{j,\sigma}p_{i,\sigma}+hc\right) +U_{d}\sum_{i}n_{d,i\downarrow}n_{d,i\uparrow}\nonumber\\&&+U_{p}\sum_{i}n_{p,i\downarrow}n_{p,i\uparrow}+E_{d}\sum_{i}n_{d,i}+E_{p}\sum_{i}n_{p,i}\nonumber\\&&+V_{dp}\sum_{\left\langle i,j\right\rangle}n_{d,i}n_{p,j}
\end{eqnarray}  
The indice $(i,\sigma)$ labels the hole Wannier state localised on the site \textbf{i} with spin $\sigma$. The operator $d_{i,\sigma}^{\dagger}$ creates a hole in the copper Wannier state $(i,\sigma)$ while the operator $d_{i,\sigma}$ destroys a hole in this state $(i,\sigma)$. The operator $p_{i,\sigma}^{\dagger}$ creates a hole in the oxygen Wannier state $(i,\sigma)$ while the operator $p_{i,\sigma}$ destroys a hole in this state $(i,\sigma)$.The operator $n_{d,i,\sigma}=d_{i,\sigma}^{\dagger}d_{i,\sigma}$ is the occupation number operator of the copper state $(i,\sigma)$ and $n_{d,i}=n_{d,i\downarrow}+n_{d,i\uparrow}$ is the holes number operator on the copper site label \textbf{i}. $n_{p,i,\sigma}$ and $n_{p,i}$ are the same operators for the oxygen states. $\left\langle i,j \right\rangle  $ indicate that the summation is performed on the nearest neighboring sites.\\
 The values of the interaction parameters $t_{dp}$, $U_{d}$, $U_{p}$, $E_{d}$, $E_{p}$ and $V_{dp}$ were estimated by differents authors \cite{Hybertsen}\cite{McMahan}\cite{Scalettar} but the results of our simulation being very sensitive these values we choose them after several tests. The values are chosen such that the low-temperature conductivity variations are large enough. The retained values of the interaction parameters  are, in eV : the hopping matrix element $t_{dp}=-1.5$, the onsite repulsive Coulomb energies $U_{d}=9$ and $U_{p}=0$, the site energies $E_{p}=3$ and $E_{d}=0$, and the Cu-O intersite Coulomb energy $V_{dp}=0.75$ Fig.\ref{fig:Emerymodel}.\\
In this model, the vacuum corresponds to the state in which all the oxygen p orbitals and copper d orbitals are fully occupied. There is one hole by elementary cell for the undoped plane.

\section{Numerical method}
\subsection{Simulation method principle}
The principle of the simulation method is detailed in references \cite{Suzuki2}\cite{Hirsch1}\cite{Hirsch2}. We recall here the essential caracteristics of the method. Within the canonical ensemble the average value of an observable, O, is given by
\begin{equation}
\label{valeurmoyen}
\left\langle O\right\rangle =tr \left( DO\right) 
\end{equation} 
where $D$ is the density operator and $Z$ the partition function
\begin{eqnarray}
\label{opdensity1}
D=\frac{e^{-\beta H}}{Z}\\
\label{partfunct1} 
Z =tr\left( e^{-\beta H}\right) 
\end{eqnarray} 
Here, $H$ is the Hamiltonian of the system and $\beta=1/k_{B}T$ the inverse temperature. In the present case, due to the presence of the off-diagonal hopping interaction, the eigenstates of the Hamiltonian, Eq.(\ref{hamiltonian1}), are not basis states, in such a way that the computation of the traces in Eqs.(\ref{valeurmoyen}) and (\ref{partfunct1}) is  problematic. The method proposed by Suzuki and Hirsch allows to get round these difficulties \cite{Suzuki1}\cite{Hirsch1}.\\
The Hamiltonian of the system is decomposed in several sub-hamiltonians $H_{r}$ comprising the creation and annihilation operators of only certain sites. Because of the fermion commutation relations some of these sub-hamiltonians do not commute. This breakup is not totally arbitrary, the sub-hamiltonians are chosen like each sub-hamiltonian cans be decomposed again in several \textit{non-overlapping} sub-system Hamiltonians $K_{r,k}$ commuting.
\begin{eqnarray}
H&=&\sum_{r=1}^{p}H_{r} \hspace{1cm} \left[ H_{r},H_{r^{\prime}}\right]\neq0\hspace{1cm}  r\neq r^{\prime}\\
\label{subhamilton1} 
H_{r}&=&\sum_{k=1}^{m_{r}}K_{r,k} \hspace{0.5cm} \left[ K_{r,k},K_{r,k^{\prime}} \right] =0 \hspace{1cm}  \forall k,k^{\prime}
\end{eqnarray} 
The previous breakups are realized with the aim of using the Trotter formula, Eq.(\ref{Trotter}), which allows to get round the non-commutativity problem of the sub-hamiltonians $H_{r}$.
\begin{equation}
\label{Trotter} 
\exp\left( -\beta\sum_{r=1}^{p}H_{r}\right) =\lim_{n\rightarrow\infty}\left[ \prod_{r=1}^{p}\left[  \exp\left( -\frac{\beta}{n}H_{r}\right) \right] \right] ^{n}
\end{equation} 
Using the Trotter formula the partition function $Z$ reads
\begin{equation}
\label{partfunct2} 
Z=\lim_{n\rightarrow\infty}Z_{n}
\end{equation} 
where $Z_{n}$ is a partition function approximant 
\begin{equation}
\label{partfunct3} 
Z_{n}=tr\left\lbrace \left[ \prod_{r=1}^{p}\left[  \exp \left( -\frac{\beta}{n}H_{r}\right) \right] \right] ^{n}\right\rbrace 
\end{equation} 
Inserting $np$ complete set of states between operators the approximant $Z_{n}$ is given by

\begin{eqnarray}
\label{partfunct4} 
 Z_{n}=\sum_{\left\lbrace \left[  \Psi_{i} \right]  \right\rbrace } \langle \Psi_{0} \vert \exp \left( -\frac{\beta}{n} H_{p}\right)  \vert \Psi_{np-1} \rangle \nonumber \\ \langle \Psi_{np-1} \vert \exp \left( -\frac{\beta}{n} H_{p-1} \right) \vert \Psi_{np-2} \rangle  \nonumber \\ \ldots \langle \Psi_{1}\vert \exp \left( -\frac{\beta}{n}H_{1}\right) \vert\Psi_{0}\rangle
\end{eqnarray} 

Here $\left[  \Psi_{i}\right] $ represents the configuration of the $np$ states $\vert\Psi_{j}\rangle$, and can be seen as the state of a $\left( d+1\right) $ dimensional classical system, $d$ being the dimension  of the studied real quantum system.  $\left\lbrace \left[  \Psi_{i} \right]  \right\rbrace $ indicates that the sum runs over all the possible configurations. This is equivalent to divide the imaginary-time interval $0\leqslant\tau\leqslant\beta$ into $n$ slices of width $\bigtriangleup\tau=\beta/n$.

An approximant of the energy of the system is
\begin{equation}
\label{energy1} 
U_{n}=-\frac{\partial}{\partial\beta}\ln Z_{n}=-\frac{1}{Z_{n}}\frac{\partial Z_{n}}{\partial \beta}
\end{equation} 

Substituting $Z_{n}$ from Eq.(\ref{partfunct4}), we obtain: 
\begin{eqnarray}
\label{energy2} 
U_{n} &=& \sum_{\left\lbrace \left[  \Psi_{i} \right]  \right\rbrace } P _{n}\left( \left[  \Psi_{i}\right]  \right) E_{n}\left( \left[ \Psi_{i}\right] \right)  \\
E_{n}\left( \left[ \Psi_{i}\right] \right) &=& \sum_{j=0}^{np-1}\frac{\langle\Psi_{j+1}\vert\frac{H_{r}}{n} \exp \left( -\frac{\beta}{n}H_{r}  \right) \vert\Psi_{j}\rangle}{\langle\Psi_{j+1}\vert \exp \left( -\frac{\beta}{n}H_{r} \right) \vert\Psi_{j} \rangle}\\
\label{Pn} 
 P_{n}\left( \left[ \Psi_{i}\right] \right)  &=& \frac{1}{Z_{n}} \langle \Psi_{0} \vert \exp \left( -\frac{\beta}{n} H_{p}\right)  \vert \Psi_{np-1} \rangle  \nonumber \\ 
& & {} \ldots \langle \Psi_{1}\vert \exp \left( -\frac{\beta}{n}H_{1}\right) \vert\Psi_{0}\rangle 
\end{eqnarray} 
The sub-hamiltonian index $r$ of each matrix element is a function of the state index $j$, so $r=1+\left( j \bmod p\right) $, $\vert\Psi_{np}\rangle=\vert\Psi_{0}\rangle$ and 
\begin{equation}
 \sum_{\left\lbrace \left[ \Psi_{i}\right] \right\rbrace }P _{n}\left( \left[  \Psi_{i}\right]  \right)=1
\end{equation} 
We can consider that each configuration $ \left[ \Psi_{i}\right]  $ of the $\left( d+1\right) $system has an energy $E_{n}\left( \left[ \Psi_{i}\right] \right)$ and a probability factor $P _{n} \left( \left[  \Psi_{i}\right]  \right)$. The computation of the average energy of the system is realized with a usual Monte Carlo method like Metropolis algorithm.\\
Using Eq.(\ref{subhamilton1}), the computed value of $U_{n}$ is:

\begin{eqnarray}
\label{Unapproxi}
U_{n} &\approx & \frac{1}{N_{p}}\sum_{\alpha=1}^{N_{p}}E_{n,\alpha} = \left\langle E_{n,\alpha}\right\rangle \\
\label{Enapproxi}  E_{n,\alpha}&=&\frac{1}{n}\sum_{i=1}^{n}\sum_{r=1}^{p}\sum_{k=1}^{m_{r}}\frac{\langle\Psi_{j+1}\vert  K_{r,k}\exp \left( -\frac{\beta}{n}\sum K_{r,l}\right) \vert\Psi_{j}\rangle}{\langle\Psi_{j+1}\vert \exp \left( -\frac{\beta}{n}\sum K_{r,l}\right) \vert\Psi_{j}\rangle} \nonumber \\ 
\end{eqnarray} 
where $N_{p}$ is the number of kept configurations, $\alpha$ is the index of configurations replacing $\left[ \Psi_{i}\right] $ and $i$ is the slice index. The indices $i, j, r$ are linked by the numbering of the $np$ states $\vert \Psi_{j}\rangle$ and verify $j=\left( i-1\right)p+r-1 $.\\
The breakup of sub-hamiltonians $H_{r}$ allows an important simplification which necessitates another approximation.
All the sub-system hamiltonians $K_{r,l}$ of the same sub-hamiltonian $H_{r}$ commute and each hamiltonian $K_{r,l}$ acts only on the state of one sub-system, thus, one writes
\begin{eqnarray}
\label{approximation1} 
\langle\Psi_{j+1}\vert \exp \left( -\frac{\beta}{n}\sum K_{r,l}\right) \vert\Psi_{j}\rangle \longleftrightarrow \nonumber \\ \prod_{l=1}^{m_{r}}\langle \varphi_{j+1,l} \vert  \exp \left( -\frac{\beta}{n}K_{r,l}\right)  \vert \varphi_{j,l}\rangle
\end{eqnarray} 
where the state $\vert \varphi_{j,l}\rangle$  is the state of the $l$ sub-system of the $r$ sub-hamiltonian of the $i$ slice. The symbol $\longleftrightarrow$ means that the left expression is replaced by the right one. This operation  means that the space of the states of  the system is considered the tensorial product of the spaces of the sub-systems states. It is not true for a fermion system, because of the state antisymmetry, the creation and annihilation operators are defined in the state space of the whole system. There is not strictly equality. Implicitly, this means that occupation numbers of the others single-particle states are not taken into account. This approximation is used by all the authors who present results obtained with world lines simulations. \\
The same factorization for the numerator of Eq. (\ref{Enapproxi}) is applied, finally, the computed values are:
\begin{eqnarray}
U_{n}^{\prime}&=&\left\langle E_{n,\alpha}^{\prime} \right\rangle \\
\label{Enalphaprime}
 E_{n,\alpha}^{\prime}&=& \frac{1}{n}\sum_{i=1}^{n}\sum_{r=1}^{p}\sum_{k=1}^{m_{r}}\frac{\langle \varphi_{j+1,k}\vert  K_{r,k}\exp \left( -\frac{\beta}{n}K_{r,k}\right) \vert \varphi_{j,k}\rangle}{\langle \varphi_{j+1,k}\vert \exp \left( -\frac{\beta}{n} K_{r,k}\right) \vert \varphi_{j,k}\rangle} \nonumber \\
\end{eqnarray} 
with
\begin{eqnarray}
\label{Pnalphaprime} 
P_{n,\alpha}^{\prime}=\frac{1}{Z^{\prime}_{n}} \prod_{i=1}^{n}   \prod_{r=1}^{p}  \prod_{k=1}^{m_{r}}\langle \varphi_{j+1,k} \vert  \exp \left( -\frac{\beta}{n}K_{r,k}\right)  \vert \varphi_{j,k}\rangle \nonumber \\ \\ 
\label{Znprime} 
Z^{\prime}_{n}=\sum \prod_{i=1}^{n}   \prod_{r=1}^{p}  \prod_{k=1}^{m_{r}}\langle \varphi_{j+1,k} \vert  \exp \left( -\frac{\beta}{n}K_{r,k}\right)  \vert \varphi_{j,k}\rangle \nonumber \\
\end{eqnarray} 
 Some probability factors $P _{n} \left( \left[  \Psi_{i}\right] \right) $ or $P_{n,\alpha}^{\prime}$ are negative: it is the ``\textit{sign problem}''. Hirsch and al. got round this difficulty by computing $\left\langle E_{n} \operatorname{sgn} P_{n} \right\rangle $ \cite{Hirsch1}. \\
The specific heat is calculated with the formula 
\begin{equation}
 c=-k \beta^{2} \frac{\partial U}{\partial\beta}
\end{equation} 
where $k$ is the Boltzman constant.\\
Using a similar method that this used for $U_{n}^{\prime}$ we obtain the fluctuation formula
\begin{equation}
\label{specifheat1} 
\frac{C}{R} \approx \beta^{2}\left\langle \left( E_{n,\alpha}^{\prime}-U_{n}^{\prime}\right) ^{2}\right\rangle 
\end{equation} 
where $C$ is the molar specific heat and $R$ the perfect gas constant.

\subsection{Simulation parameters}
\begin{figure}
\begin{center}
\includegraphics[width=7cm]{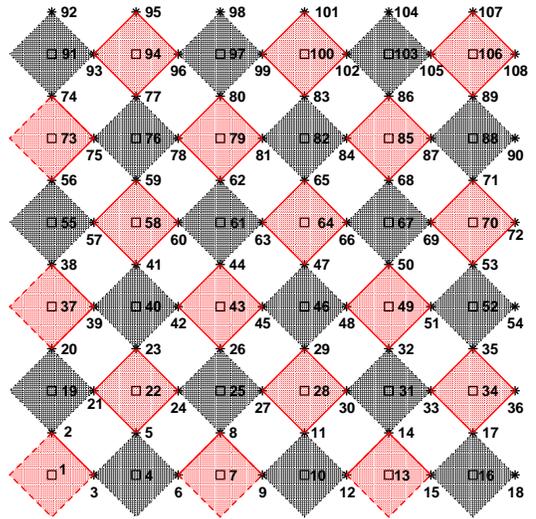}
\end{center}
\caption{(Color online). The breakup of the CuO$_{2}$ plane into two sub-hamiltonians with periodic boundary condition. The grey (or red) sub-systems belong to the sub-hamiltonian 1, and the dark sub-systems belong to the sub-hamiltonian 2.}
\label{fig:Res2D}
\end{figure}

Our simulation model of the CuO$_{2}$ plane contains $6\times6$ elementary cells with periodic boundary conditions. Fig. \ref{fig:Res2D} shows the breakup of the model. There is one sub-system around each copper site. All sub-systems are identical and are composed by one copper site and four oxygen sites. These sub-systems are grouped together in two sub-hamiltonians $\left( p=2\right) $ in such a way that the sub-systems of one sub-hamiltonian do not have common site. The sub-system hamiltonians are identical and called $K$. Using the numbering 1 of the sub-system sites given in the figure \ref{fig:elematrix-2}, the sub-system hamiltonian reads
\begin{eqnarray}
\label{hamiltonianK} 
 K=&&t_{dp}\sum_{\sigma}\left( d^{\dagger}_{3\sigma}\left( p_{1\sigma}+p_{2\sigma}+p_{4\sigma}+p_{5\sigma}\right) +hc\right)\nonumber\\
&&+U_{d}\: n_{d3\downarrow}n_{d3\uparrow}\nonumber\\
&&+\frac{U_{p}}{2}\left( n_{p1\downarrow}n_{p1\uparrow}+n_{p2\downarrow}n_{p2\uparrow}+ n_{p4\downarrow}n_{p4\uparrow}+n_{p5\downarrow}n_{p5\uparrow}\right)\nonumber\\
&&+E_{d}\: n_{d3}+\frac{E_{p}}{2}\left( n_{p1}+n_{p2}+ n_{p4}+n_{p5}\right)\nonumber\\
&&+V_{dp}\: n_{d3}\left(n_{p1}+n_{p2}+n_{p4}+ n_{p5}\right)  
\end{eqnarray} 
The factor $1/2$ is introduced to take into account that each oxygen site, numbered 1,2,4 and 5, belongs to two sub-systems.\\
The state space dimension of each sub-system being $2^{10}$ the method necessitates the diagonalization of only one $\left( 1024\times1024\right) $ matrix. A constant value is added to the eigenvalues in such a way that the ground level is zero.\\
The imaginary-time interval is divided into twenty four slices $\left( n=24\right) $. Thus the (2+1)dimensional system is composed of forty eight 2D-lattices $\left( np=48\right) $.\\
Each simulation begins with thermalization. Then decreasing temperature (100 points) is programmed, at each temperature point the averages are computed on $100$ configurations of the 3D-array. There is about 50,000 Monte Carlo steps between two kept configurations.\\
Simulations were performed for different dopings $\delta$. For $6\times6$ elementary cells, there are 36 holes $\left( 18\uparrow+18\downarrow\right) $ for doping $\delta=0$ and 72 holes $\left( 36\uparrow+36\downarrow\right) $ for $\delta=1$.\\
The energy average $U=\langle E_{i} \rangle$ of the model is calculated and the ratio $C/R$ is computed with the energy fluctuations formula,  Eq. (\ref{specifheat1}).\\
 The comparison of $C/R$ with the derivative of the cell energy can be used as criterion of the numerical method quality.\\

\subsection{Influence of the state numbering}
\label{sec:example}
\begin{figure}
\begin{center}
\includegraphics[width=6cm]{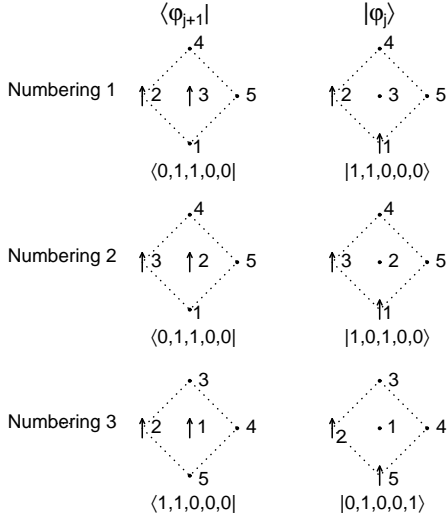}
\end{center}
\caption{Three different numberings of the sub-system sites.}
\label{fig:elematrix-2}
\end{figure}

Considering that the sub-system hamiltonian $K$ does not modify the particle numbers, the matrix of the hamiltonian $K$ cans be decomposed in sub-matrices corresponding to sub-spaces of fixed particle numbers. The maximum dimension of these sub-spaces is 100. It corresponds to the four sub-spaces of 2 spins up and 2 spins down, or 2 spins up and 3 spins, or 3 spins up and 2 spins down, or 3 spins up and 3 spins down. The dimension of the sub-space corresponding to 2 spins up and zero spin down is $C_{5}^{2}=10$. This sub-space is chosen to test the influence of the single-particle state numbering on the matrix of the hamiltonian $K$  and consequently on the probability factor sign.\\
The numbering principle is chosen such that the numbers of the single-particle states correspond to the sites numbers in the present case where there is only spins up. An  off-diagonal matrix elements $\langle\varphi_{j+1}\vert K \vert\varphi_{j}\rangle$ is calculated for three different numberings of the sites. The figure \ref{fig:elematrix-2} shows the corresponding situations. The element values are

\begin{equation}
\begin{split}
\textrm{numbering 1}& \\  \langle\varphi_{j+1}\vert K \vert\varphi_{j}\rangle  &=\langle0,1,1,0,0\vert K \vert 1,1,0,0,0\rangle=-t_{dp} 
\end{split}
\end{equation} 

\begin{equation}
\begin{split}
\textrm{numbering 2}& \\ \langle\varphi_{j+1}\vert K \vert\varphi_{j}\rangle  &=\langle0,1,1,0,0\vert K \vert 1,0,1,0,0\rangle=t_{dp} 
\end{split}
\end{equation} 

\begin{equation}
\begin{split}
\textrm{numbering 3}& \\ \langle\varphi_{j+1}\vert K \vert\varphi_{j}\rangle  &=\langle1,1,0,0,0\vert K \vert 0,1,0,0,1\rangle=-t_{dp} 
\end{split}
\end{equation} 

\begin{figure}
\begin{center}
\includegraphics[width=8cm]{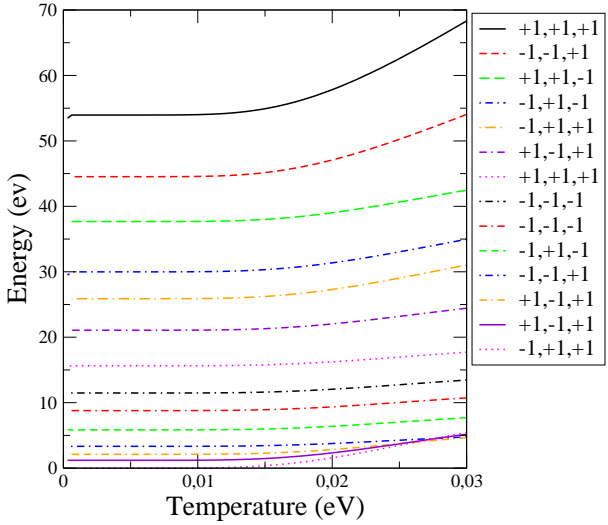}
\end{center}
\caption{(Color online). Energies of 3D array configurations versus temperature. The probability factor signs correspond respectively to the three numberings of the figure \ref{fig:elematrix-2}.}
\label{fig:energsigne-1}
\end{figure}

The sign of the off-diagonal matrix elements are dependent of the state numbering. Indeed the change-of-basis matrices are unitary (orthogonal) matrices and the elements are 0 or $\pm1$ with only one $+1$ or one $-1$ by row and column, the others elements being $0$. They are permutation matrices except that some matrix elements are $-1$ instead $+1$ because of the antisymmetry.\\
These remarks presented for the particular $\left( 10\times10\right) $ sub-matrix hold for the all sub-matrices of the hamiltonian $K$.
For a numbering change, the change-of-basis matrices (and the inverse matrices) act on the sub-matrices of the operators $\exp \left( -\beta K \right)$  and $K \exp \left( -\beta K \right)$. Thus the sign of some elements of these sub-matrices change when as the absolute values are equal. For a fixed particle number sub-space, the modified elements of the two matrices are the same. Thus the sign of each ratio of the sum in the relation (\ref{Enalphaprime}) is not change, whereas the sign of some factors of the product in the  relation (\ref{Pnalphaprime}) change.\\
These remarks show that the energy $E_{n,\alpha}^{\prime}$,  Eq. (\ref{Enalphaprime}), of each configuration of the 3D array is independent of the numbering while the sign of the probability factor $P_{n,\alpha}^{\prime}$, Eq. (\ref{Pnalphaprime}), cans change. The figure \ref{fig:energsigne-1} shows the variations of the energies $E_{n,\alpha}^{\prime} $ of different 3D array configurations with the temperature, and the probability factor signs, for the three numberings of the figure \ref{fig:elematrix-2}.

\subsection{The probability factor signs can be ignored}
\label{sec:signproblem}
The partition fonction $Z_{n}^{\prime}$, Eq. (\ref{Znprime}), and the probability factor $P_{n,\alpha}^{\prime}$, Eq. (\ref{Pnalphaprime}), can be written \\
 
\begin{eqnarray}
Z_{n}^{\prime}&=&\sum a_{\alpha}\\
P_{n,\alpha}^{\prime}&=&\frac{a_{\alpha}}{Z_{n}^{\prime}} \\
a_{\alpha}&=&\prod_{i=1}^{n}   \prod_{r=1}^{p}  \prod_{k=1}^{m_{r}}\langle \varphi_{j+1,k} \vert  \exp \left( -\frac{\beta}{n}K_{r,k}\right)  \vert \varphi_{j,k}\rangle \nonumber \\
\end{eqnarray} 
where $a_{\alpha}$ is relative to 3D array configuration indiced $\alpha$. The energy $E_{n,\alpha}^{\prime}$ and the absolute value of $a_{\alpha}$ are properties of the 3D array configuration, they depend only on the configuration whereas the sign of $a_{\alpha}$ is dependent of the state numbering.\\
 During numbering change, some positive terms $a_{\alpha}$ become negative and some negative terms become positive. The predictions of the method must be independent of the state numbering, the partition funcion $Z_{n}^{\prime}$ and the energy average must be unchanged, thus

\begin{eqnarray}
 \sum_{\left\lbrace \alpha_{p \rightarrow n} \right\rbrace }  a_{\alpha} =-\sum_{\left\lbrace \alpha_{n \rightarrow p} \right\rbrace } a_{\alpha} =\sum_{\left\lbrace \alpha_{n \rightarrow p} \right\rbrace } \vert a_{\alpha}\vert\\
 \sum_{\left\lbrace \alpha_{p \rightarrow n} \right\rbrace } a_{\alpha}  E_{\alpha}=-\sum_{\left\lbrace \alpha_{n \rightarrow p} \right\rbrace } a_{\alpha} E_{\alpha} =\sum_{\left\lbrace \alpha_{n \rightarrow p} \right\rbrace } \vert a_{\alpha}\vert E_{\alpha} 
\end{eqnarray}
where $E_{\alpha}$ replaces $E_{n,\alpha}^{\prime}$, $\left\lbrace \alpha_{n \rightarrow p } \right\rbrace $ represents the set of 3D array configurations of which the terms $a_{\alpha}$ become positive and $\left\lbrace \alpha_{p \rightarrow n} \right\rbrace $ the set of 3D array configurations of which the terms $a_{\alpha}$ become negative. The energy averages of these sets of configurations are equal:

\begin{equation}
\frac{\sum_{\left\lbrace \alpha_{p \rightarrow n} \right\rbrace } a_{\alpha}  E_{\alpha}}{\sum_{\left\lbrace \alpha_{p \rightarrow n} \right\rbrace }  a_{\alpha} }=\frac{\sum_{\left\lbrace \alpha_{n \rightarrow p} \right\rbrace } a_{\alpha}  E_{\alpha}}{\sum_{\left\lbrace \alpha_{n \rightarrow p} \right\rbrace }  a_{\alpha}}
\end{equation} 
This result is true for all the different possible numberings of the single-particle states of the sub-system.\\
 The factor $a_{\alpha}$ of each configuration $\alpha$ is positive or negative depending on the numbering and the energy of each 3D array configuration is a contributation to averages computed with positive and negative factors. Thus we can deduce that, for a given numbering, the energy averages computed separately for the positive en negative probability factors are equal:
\begin{equation}
\frac{\sum_{\left\lbrace \alpha_{>0} \right\rbrace } a_{\alpha} E_{\alpha}}{\sum_{\left\lbrace \alpha_{>0} \right\rbrace }  a_{\alpha} }=\frac{\sum_{\left\lbrace \alpha_{<0} \right\rbrace } a_{\alpha}  E_{\alpha}}{\sum_{\left\lbrace \alpha_{<0} \right\rbrace }  a_{\alpha} } 
\end{equation}  
where $\left\lbrace \alpha_{>0} \right\rbrace $ and $\left\lbrace \alpha_{<0} \right\rbrace $ represent respectively the sets of 3D array configurations with positive and negative probability factors. This equality between the energy averages computed separately with the ``\textit{negative}'' and ``\textit{positive}'' configurations was already remarked \cite{Linden}. Ours computational results confirm this equality. This behaviour is independent of the number of kept configurations, the number of slices, the size of the model and the temperature.\\
Using a well-known arithmetic result one cans write

\begin{eqnarray}
& & \frac{\sum_{\left\lbrace \alpha_{>0} \right\rbrace } a_{\alpha} E_{\alpha}}{\sum_{\left\lbrace \alpha_{>0} \right\rbrace }  a_{\alpha} }=\frac{\sum_{\left\lbrace \alpha_{<0} \right\rbrace } a_{\alpha}  E_{\alpha}}{\sum_{\left\lbrace \alpha_{<0} \right\rbrace }  a_{\alpha} } \nonumber \\& =&\frac{\sum_{\left\lbrace \alpha_{>0} \right\rbrace } a_{\alpha} E_{\alpha} -\sum_{\left\lbrace \alpha_{<0} \right\rbrace } a_{\alpha}  E_{\alpha}}{\sum_{\left\lbrace \alpha_{>0} \right\rbrace }  a_{\alpha} -\sum_{\left\lbrace \alpha_{<0} \right\rbrace }  a_{\alpha}} \nonumber \\ &=&\frac{\sum_{\left\lbrace \alpha \right\rbrace } \vert a_{\alpha}\vert E_{\alpha} }{\sum_{\left\lbrace \alpha \right\rbrace } \vert a_{\alpha}\vert } \label{somplus}  \\& =&\frac{\sum_{\left\lbrace \alpha_{>0} \right\rbrace } a_{\alpha} E_{\alpha} +\sum_{\left\lbrace \alpha_{<0} \right\rbrace } a_{\alpha}  E_{\alpha}}{\sum_{\left\lbrace \alpha_{>0} \right\rbrace }  a_{\alpha} +\sum_{\left\lbrace \alpha_{<0} \right\rbrace }  a_{\alpha}} \nonumber \\& =&\frac{\sum_{\left\lbrace \alpha \right\rbrace } a_{\alpha} E_{\alpha}} {\sum_{\left\lbrace \alpha \right\rbrace }  a_{\alpha}}=\sum _{\left\lbrace \alpha \right\rbrace } P_{\alpha} E_{\alpha} \label{somoins} 
\end{eqnarray} 
where $P_{\alpha}$ replaces $ P_{n,\alpha}^{\prime}$. The equality of terms (\ref{somplus}) and (\ref{somoins}) indicates that the sign of the probability factor cans be ignored. We use this remark for our simulations. The computational results are actually independent of the single-particle state numbering.

\section{Test of the validity of the method}

The validity of the method is tested with the two-dimensional Hubbard model on a square lattice at half filling. This model was extensively studied with different methods \cite{Paiva}\cite{Duffy}\cite{Scalettar2}. The model and simulation parameters were chosen to compare our results with the results of ref. \cite{Paiva} (fig. 6 and fig. 8). The Hubbard hamiltonian is 
\begin{equation}
\label{hamiltonian2} 
H =t\sum_{\left\langle i,j\right\rangle,\sigma }\left( c^{\dagger}_{j,\sigma}c_{i,\sigma}+hc\right) +U\sum_{i}n_{i\downarrow}n_{i\uparrow}
\end{equation}
The imaginary-time interval is divided into eight slices $\left( n=8\right) $, the lattice is $6\times6$ with periodic boundary conditions, with eighteen spins up and eighteen spins down ($18\uparrow+18\downarrow$). The interaction parameters are $t=-1$ and $U=2$ or $U=10$. All sub-systems are identical and are composed by four sites. These sub-systems are grouped together in two sub-hamiltonians $\left( p=2\right) $.\\
\begin{figure}
 \begin{center}
\includegraphics[width=7cm]{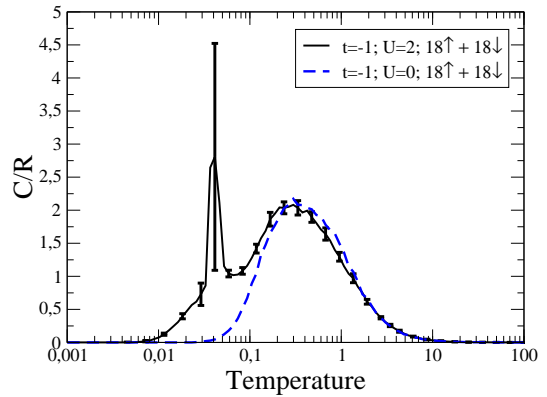}
 \end{center}
\caption{(Color online) Molar specific heat versus temperature for the Hubbard model at half filling. $t=-1$, $U=2$ and the lattice is $6\times6$. The dashed line curves (blue) correspond to $U=0$.}
\label{fig:energspecHub-1}
\end{figure}
 The Fig. \ref{fig:energspecHub-1} displays the molar specific heat curve as a function of  temperature $kT$ for $U=2$. The curve is an average of eighteen simulations while the curve for $U=0$ is an average of six simulations. There are 100 temperature points by curve (20 points by decade).  For the sake of clarity all the error bars are not shown. These curves are very similar to the curves in ref. \cite{Paiva}. In particular, a two peaks structure is observed, with a broad high temperature peak and a narrower peak at low temperature. However, the temperature range is translated by a factor four toward the low temperature and there is a ratio four between the $C/R$ amplitudes. This ratio is consistent with the definition $C=\partial U/\partial T$. Indeed, in ref. \cite{Paiva} the slice number $n$ is not constant while $n=8$ in our simulations, this difference can explain the temperature translation. The intensity of the low temperature peak of our simulation is larger than in ref. \cite{Paiva}. Note that the error bars are large in the temperature range of the low temperature peak. This indicates that there are large fluctuations in this temperature range (0.033-0.05). A simulation with 200 temperature points in the range (0.01-0.1) was performed to understand this problem. Fig \ref{fig:energHub-1} shows the energy curves in this temperature range. The dashed line curve with error bar corresponds to the average of eighteen simulations with 100 points in the temperature range (0.001-100). The solid line curve is the result of the simulation with 200 points in the temperature range (0.01-0.1). The energy of the states generated into the temperature range (0.033-0.05), oscillate between two values. The error bars are very small outside this temperature range. This indicates that a first order transition occurs. This explains the large fluctuations of the low temperature specific heat peak. The results in ref. \cite{Paiva} are obtained with the determinantal QMC simulations, based also on the Trotter formula. The specific heat curves $C\left( T\right) $ are evaluated by differentiation of the energy data with about 10 temperature points by decade. Our specific heat results are raw data of the simulations, computed with fluctuation formula. These differences of methods and numbers of points explain the differences between the results.\\

\begin{figure}
 \begin{center}
\includegraphics[width=7cm]{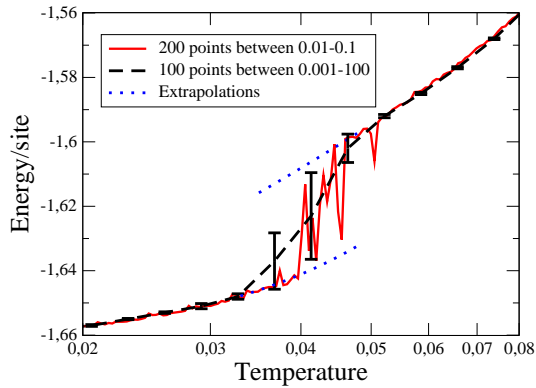}
 \end{center}
\caption{(Color online) Site energy versus temperature for the Hubbard model at half filling. $t=-1$, $U=2$ and the lattice is $6\times6$. The dashed line curve (black) corresponds to the  average of 18 simulations with 100 points. The error bars are very small outside the transition range. The solid line curve (red) is obtained with 200 points in temperature range (0.01-0.1). The dot line curves (blue) are extrapolations of the energy curves. }
\label{fig:energHub-1}
\end{figure}

\begin{figure}
 \begin{center}
\includegraphics[width=7cm]{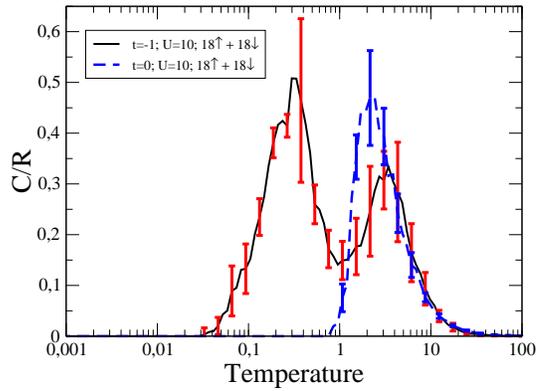}
 \end{center}
\caption{(Color online) Molar specific heat versus temperature for the Hubbard model at half filling. $t=-1$, $U=10$ and the lattice is $6\times6$. The dashed line curves (blue) correspond to $t=0$.}
\label{fig:energspecHub-2}
\end{figure}

 The Fig. \ref{fig:energspecHub-2} displays the molar specific heat curve for $U=10$ (average over 18 simulations). This curve is very similar to the curve in ref. \cite{Paiva}. The intensity of the peak for $t=0$ is slightly higher than the high temperature peak but the uncertainty can explain this difference. For $t=0$ the sub-system hamiltonian is diagonal and there is no sign problem. The error bars indicate that, for $t=-1$ and $U=10$, the algorithm generates states which stay a long time in local minima for increasing and decreasing temperatures. This problem is less important for the CuO$_{2}$ plane. Thus a small number of simulations is required to obtained small error bars.\\

\begin{figure}
 \begin{center}
\includegraphics[width=7cm]{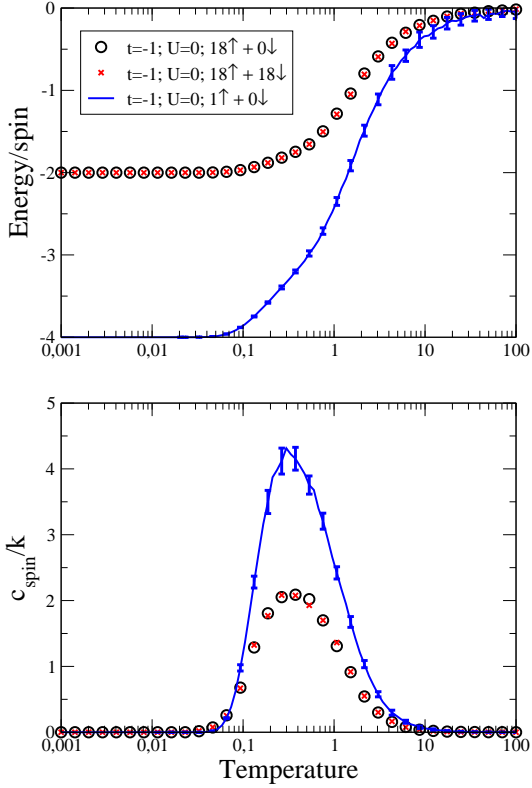}
 \end{center}
\caption{(Color online) Spin energy and spin specific heat versus temperature for the Hubbard model for different fillings. $t=-1$, $U=0$ and the lattice is $6\times6$. The solid line curves (blue) correspond to one spin up on the array.}
\label{fig:energspecHub-3} 
\end{figure}

The test of the validity of the method is completed with simulations for $U=0$ (fig.\ref{fig:energspecHub-3}). For $U=0$, the half filling ($18\uparrow+18\downarrow$) corresponds to two no-correlated quantum gases with the same energy and specific heat. The simulations for $18\uparrow$ and no spin down give identical results for the spin energy and the spin specific heat. The ground state energy by spin is $2t$. For only one spin ($1\uparrow$) on the array the antisymmetry of states is not relevant, thus there is no sign problem. The spin energy and specific heat for one spin are twice the values computed for $18\uparrow+18\downarrow$ or $18\uparrow+0\downarrow$. As expected, the ground state energy is $4t$. The value of the ratios of spin energy and specific heat computed for $1\uparrow+0\downarrow$ and $18\uparrow+18\downarrow$ indicate that the sign problem is correctly got round in our simulations.\\
The similarity between the results of ref. \cite{Paiva} and our results, and the good overall agreement between simulation results with no sign problem and simulation results with sign problem confirm that the sign is not an actual problem for our simulations. The simulation method is valid.

\section{CuO$_{2}$  plane: First results}
\begin{figure}
 \begin{center}
\includegraphics[width=7cm]{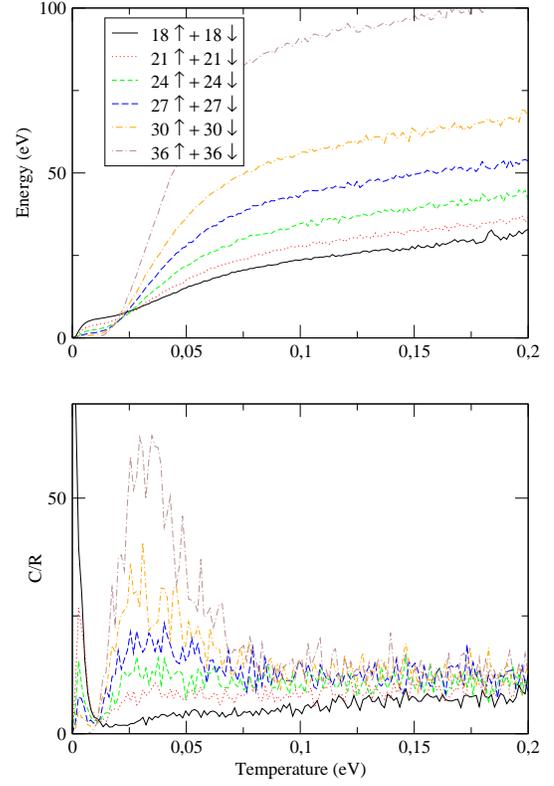}
 \end{center}
 \caption{(Color online). Energy and specific heat versus temperature for different dopings of the CuO$_{2}$ plane.}
 \label{fig:energspec1}
\end{figure}

The Figs. \ref{fig:energspec1} and \ref{fig:energspec2} display the energy and specific heat curves versus temperature $kT$ (in $eV$) for the CuO$_{2}$  plane. The curves of Fig. \ref{fig:energspec2} are the averages of five simulations. Whatever the algorithm quality, one notices that the low temperature state of the model is the ground state. We observe a crossing-point on the set of specific heat curves with $C/R\sim2.7$. This phenomenon is often observed for in correlated systems\cite{Vollhardt}. Fig. \ref{fig:res18-1} shows one of the 48 2D-lattice states of the 3D-array  for doping $\delta=0$, while Fig. \ref{fig:res24-1} shows one 2D-lattice state for $\delta=0.33$ $\left( 24\uparrow+24\downarrow\right) $. These 3D-arrays are the last generated at the lowest temperature ($kT\simeq2.5\,10^{-4}$). For $\delta=0$, we remark some antiferromagnetic loops, and antiferromagnetic chains for $\delta=0.33$. Only the oxygen sites are occupied in AF loops. It is the same for AF chains except that the end chain sites can be copper sites. The loops can be related to the circulating current proposed by C. M. Varma \cite{Varma1},\cite{Varma2},\cite{Varma3},\cite{Varma4},\cite{Sidis},\cite{Normand} while the chains recall the stripes proposed by several authors \cite{Carlson}\cite{Kivelson}. This comforts us to study the average temperature dependence of the chain and loop numbers.  Fig. \ref{fig:chainloop1} shows these temperature dependence. It appears clearly that, for $\delta=0$, the $C/R$ peak  is associated with the chain and loop numbers temperature dependence. There is no evidence for the other dopings. \\

\begin{figure}
 \begin{center}
\includegraphics[width=7cm]{figure-6.eps}
 \end{center}
 \caption{(Color online). Energy and specific heat versus temperature for different dopings.}
 \label{fig:energspec2}
\end{figure}

\begin{figure}
\begin{center}
\includegraphics[width=6cm]{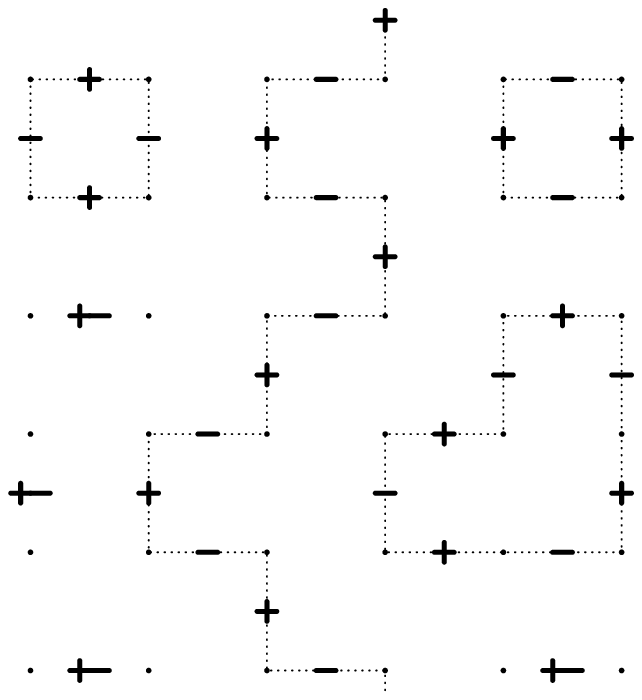}
\end{center}
\caption{Typical 2D-lattice for $\delta=0$ at $kT=2.5 \, 10^{-4}\,eV$. Points indicate non-occupied copper sites. Sign + indicates spin up and sign - spin down. The dot lines show the antiferromagnetic loops.}
\label{fig:res18-1}
\end{figure}

\begin{figure}
\begin{center}
\includegraphics[width=6cm]{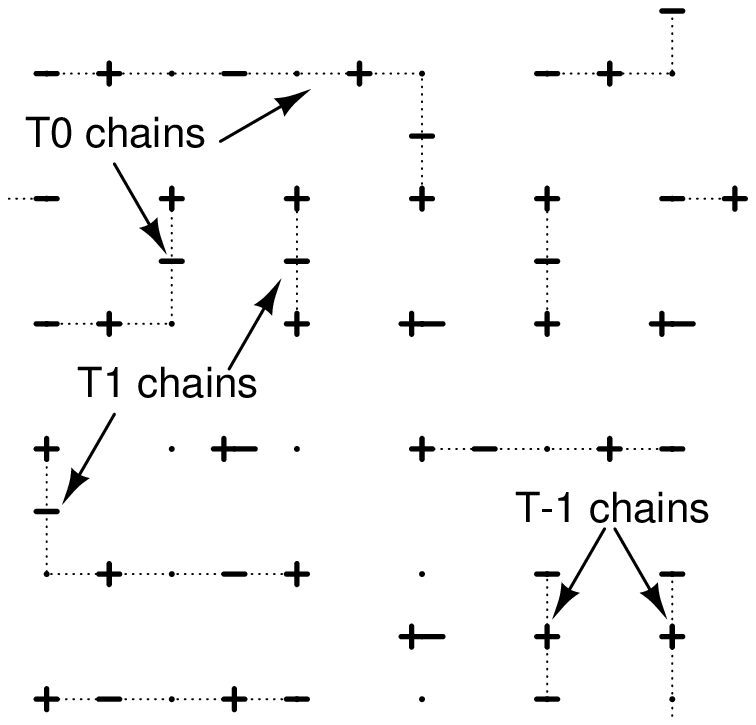}
\end{center}
\caption{Typical 2D-lattice for $\delta=0.33$ at $kT=2.5 \, 10^{-4}\,eV$. Points indicate non-occupied copper sites. Sign + indicates spin up and sign - spin down. The dot lines show the antiferromagnetic chains.}
\label{fig:res24-1}
\end{figure}

\begin{figure}
 \begin{center}
\includegraphics[width=7cm]{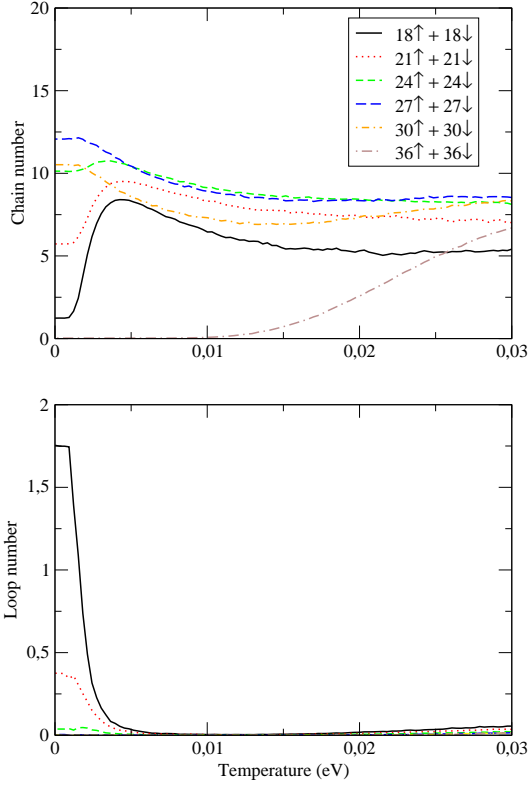}  
 \end{center}
 \caption{(Color online). Chain and loop numbers versus temperature for different dopings.}
 \label{fig:chainloop1}
\end{figure}

\begin{figure}
\begin{center}
\includegraphics[width=6cm]{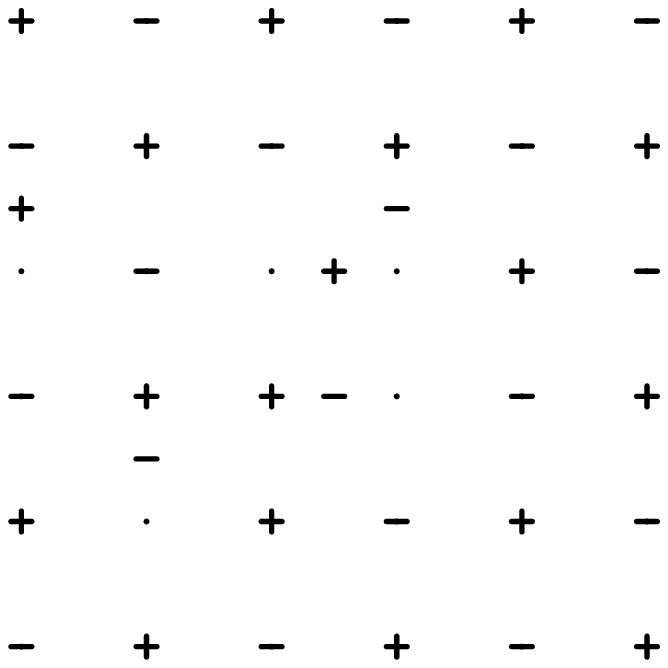}
\end{center}
\caption{Typical 2D-lattice for $\delta=0$ at $kT\simeq0.025\,eV$. Points indicate non-occupied copper sites. Sign + indicates spin up and sign - spin down.}
\label{fig:res18-2}
\end{figure}

\begin{figure}
\begin{center}
\includegraphics[width=6cm]{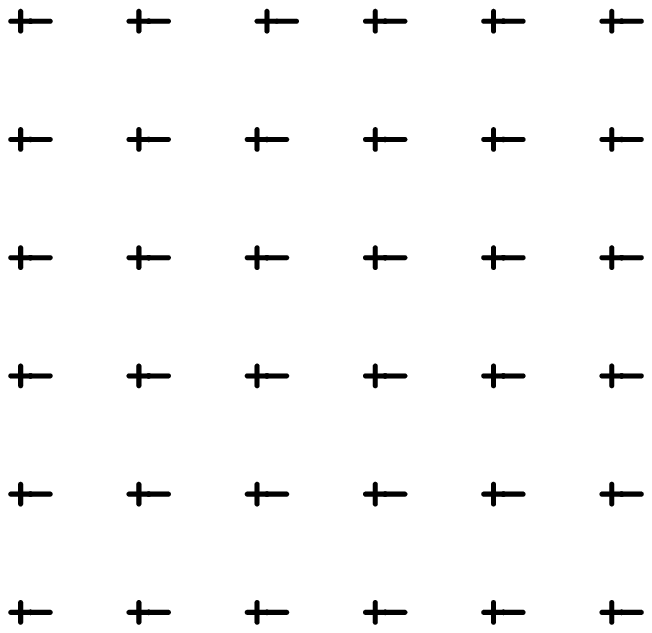}
\end{center}
\caption{Typical 2D-lattice for $\delta=1$ at $kT=10^{-5}\,eV$. Sign + indicates spin up and sign - spin down.}
\label{fig:res36-1}
\end{figure}

\begin{figure}
 \begin{center}
\includegraphics[width=7cm]{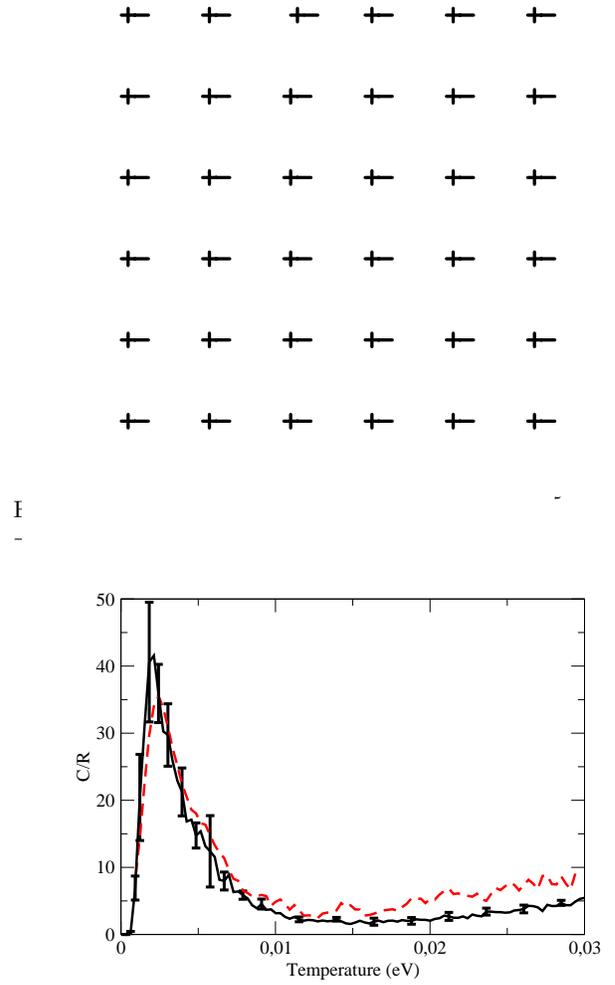}  
 \end{center}
\caption{(Color online). Comparaison of C/R curves for $\delta=0$. Solid line curve corresponds to fluctuations computation, dashed line curve is obtained by energy derivative.}
 \label{fig:comspecheat-1}
\end{figure}

Fig. \ref{fig:res18-2} displays one 2D-lattice at temperature $kT\simeq0.025$ for $\delta=0$. A tendency to antiferromagnetic configuration, where only copper sites are occupied, is observed.\\
For $\delta=1$ all the low temperature 2D-lattices are identical to the 2D-lattice of Fig. \ref{fig:res36-1}.\\
The numerical method quality is tested by comparing the specific heat computed with Eq. (\ref{specifheat1})  with the derivative of the energy curve for doping $\delta=0$. Running average is applied to the energy curve before derivative. The derivative must be divided by the number of elementary cells to obtain $C/R$. The Fig. \ref{fig:comspecheat-1} shows the curves obtained with the two methods. There is a good agreement for low temperatures.\\
\section{Conductivity}
\label{sec:conductivity}
\begin{figure}
 \begin{center}
\includegraphics[width=7cm]{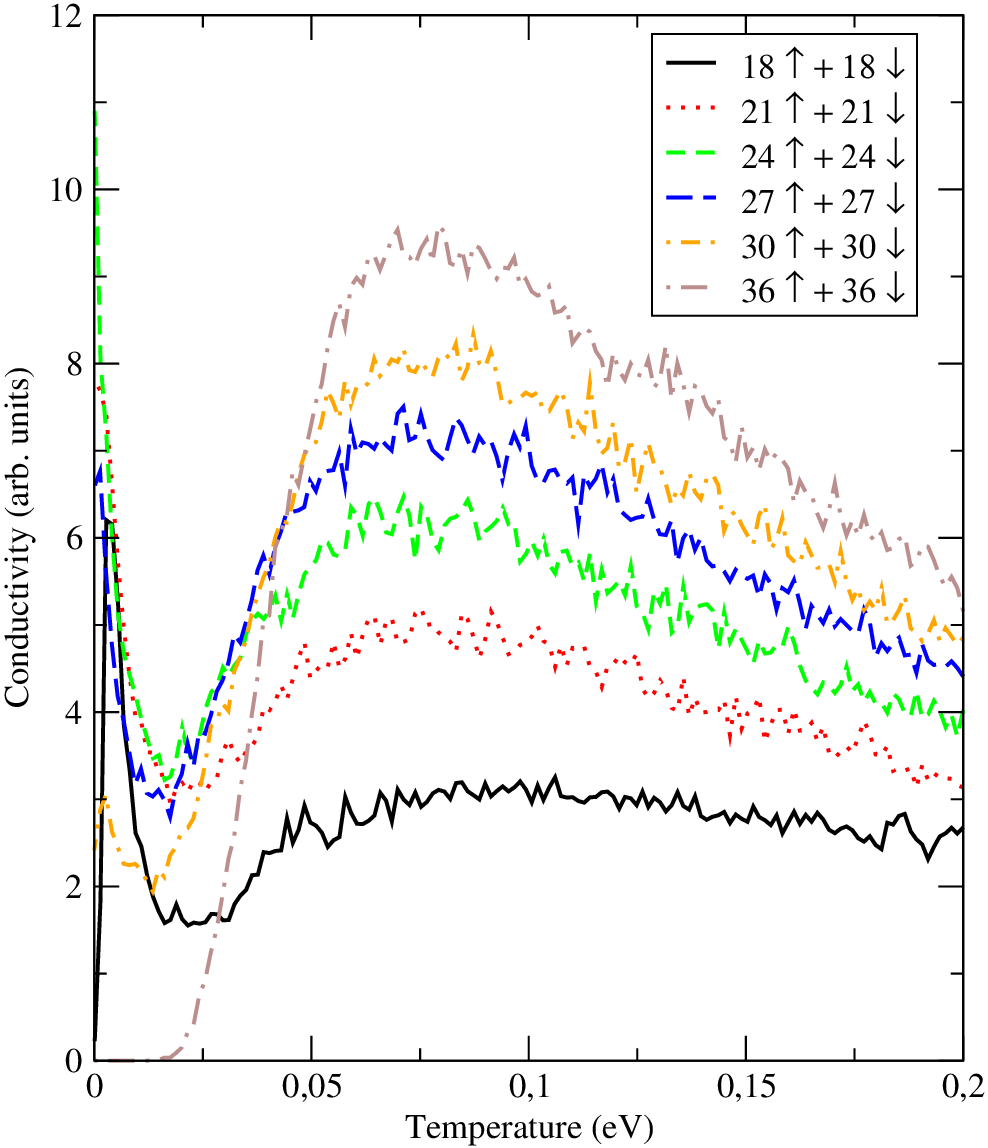}  
 \end{center}
 \caption{(Color online). Conductivity versus temperature for different dopings.}
 \label{fig:conductiv-1}
\end{figure}

\begin{figure}
 \begin{center}
\includegraphics[width=7cm]{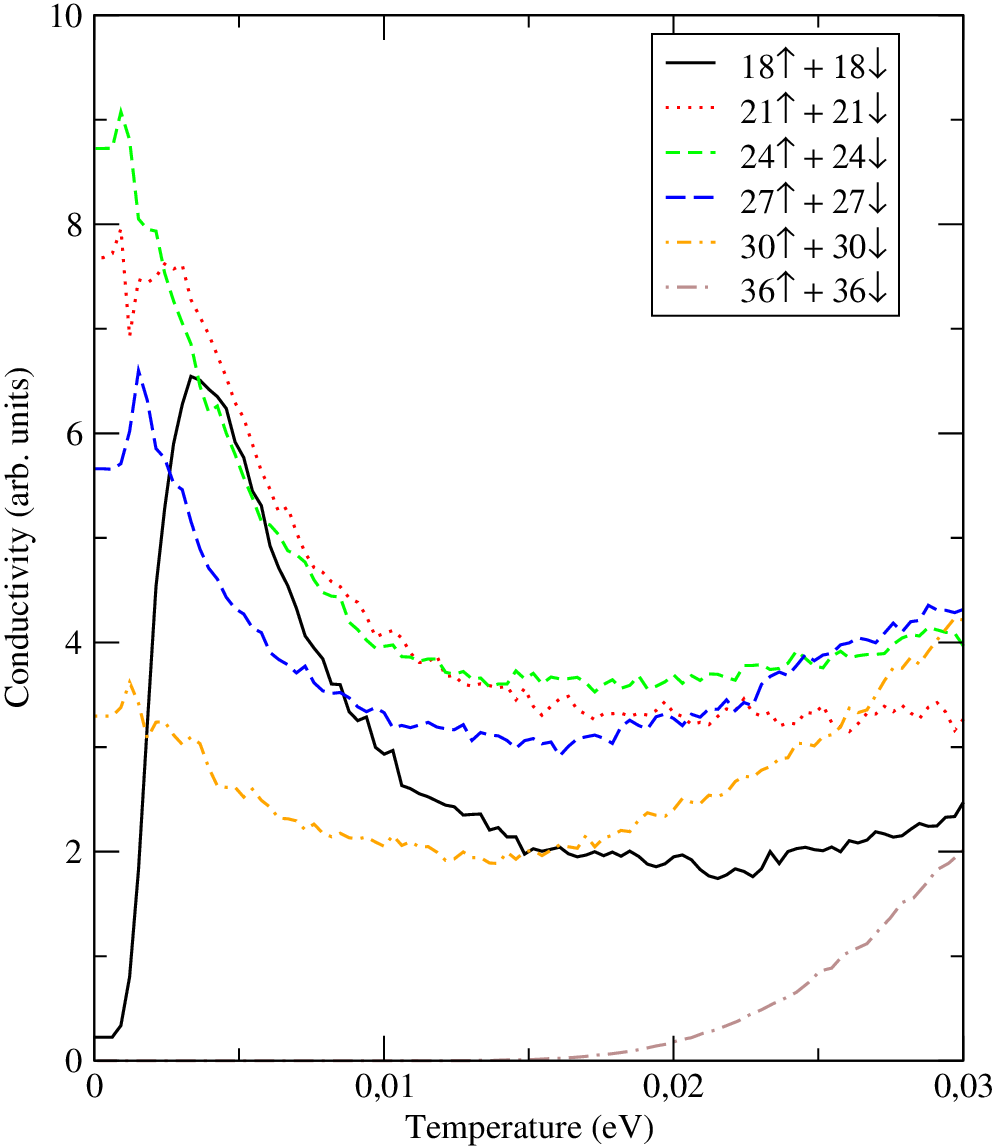}  
 \end{center}
 \caption{(Color online). Conductivity versus temperature for different dopings.}
 \label{fig:conductiv-2}
\end{figure}

The conductivity of the model for the different doping is an important question. The present simulation method does not allow rigorous conductivity computation because of current-density operator defined by Scalapino and al. \cite{Scalapino1} is not diagonal and can not break up into sub-system operator \cite{Hirsch2}. To obtain rough approximation of the current-density at imaginary-time $\tau=\frac{\beta}{np}m$, we calculated:
\begin{eqnarray}
 j_{x}\left( l;m\right)=&&\sum_{\sigma}n_{\sigma,l;m}\left( 1-n_{\sigma,l^{\prime};m}\right)  n_{\sigma,l^{\prime};m+p} \nonumber \\ &&\left( 1-n_{\sigma,l;m+p}\right)-n_{\sigma,l^{\prime};m}\left( 1-n_{\sigma,l;m}\right)\nonumber\\&& n_{\sigma,l;m+p}\left( 1-n_{\sigma,l^{\prime};m+p}\right) 
\end{eqnarray}
The coordinates of the $l$ and $l^{\prime}$ sites  verify $x_{l^{\prime}}=x_{l}+1$  and $y_{l^{\prime}}=y_{l}$. $n_{\sigma,l;m}$ is the occupation number of the single-particule state $\vert \sigma,l\rangle$ of the 2D-lattice numbered $m$.
This expression reads 
\begin{equation}
\langle\psi_{m+p}\vert \sum_{\sigma}c\dagger _{\sigma,l^{\prime}}c_{\sigma,l}-c\dagger _{\sigma,l}c_{\sigma,l\prime}\vert \psi_{m}\rangle 
\end{equation} 
where the state $\vert\psi_{m}\rangle$ corresponds to the state of $l$ and $l^{\prime}$ sites. The computed conductivity is: 
\begin{equation}
 \Lambda_{xx}\left( m\right) =\frac{1}{np}\sum_{i=1}^{np} \sum_{l=1}^{N_{s}} j_{x}\left( l;m+i\right)j_{x}\left( l;i\right)
\end{equation} 
where $N_{s}$ is the sites number. The DC conductivity is obtained after a discrete Fourier transform.
Figs. \ref{fig:conductiv-1} and \ref{fig:conductiv-2} show the DC conductivity for the different dopings. If this calculation is relevant, we observe two metal-insulator transitions for $\delta=0$ and $\delta=1$.\\

\section{Discussion}
\label{sec:discussion}
\subsection{Doping $\delta=0$}
By contrast with the behaviour usually expected, the low-temperature state is not antiferromagnetic (AF) with only occupied copper sites, this can be observed on the figure \ref{fig:res18-1}. As the temperature is decreased, the system state seems to go to AF state (Fig. \ref{fig:res18-2}) and the conductivity decreases but it rapidly increases  before dropping to zero. The ground state is essentially composed with  antiferromagnetic loops. Apparently this ground state is a new insulator phase which is not a Mott insulator as the AF state. The sharp conductivity increasing is difficult to explain, while the conductivity dropping corresponds to the transformation of chains into loops. Indeed the temperature dependence of the chains and loops numbers have similar dependencies that the conductivity. The specific heat peak corresponds to this transformation.
\subsection{Other dopings}
\label{sec:otherdopings}
\subsubsection{Conductivity and chains}
\begin{figure}
 \begin{center}
\includegraphics[width=7cm]{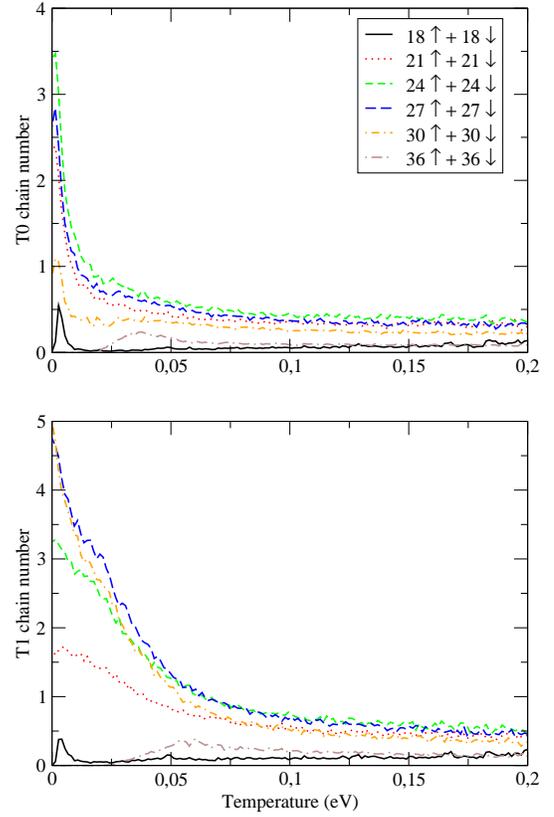}  
 \end{center}
 \caption{(Color online). T0 and T1 chain numbers versus temperature for different dopings.}
 \label{fig:triplet-1}
\end{figure}

The conductivity curves for $\delta=0.16$, $\delta=0.33$ and $\delta=0.5$ show that the system exhibits a metallic behaviour at high and low temperatures with a semiconductor like behaviour for the intermediate range. Thus, these two metallic behaviours must be different. At low temperature, the conductivity shows a maximum for $\delta=0.33$. The low temperature metallic behaviour seems to be associated with the chain number but the temperature dependence of the conductivity and chain number curves are not similar. A more detailed analysis of the chains is needed.\\
  There are three different types of chains: chains without ``order'', antiferromagnetic chains and antiferromagnetic chains with the two end sites being copper sites. The last chain type is composed of three kinds of chains according to whether the spins of the holes of the chain ends are identical or opposite. The numbers of this different chains are computed. Fig. \ref{fig:triplet-1} shows the curve of the number of chains with opposite end spins. These chains, called T0, have an even holes number. Fig. \ref{fig:triplet-1} displays also the curve of number of chains with spin up ends, which have an odd hole number. These chains are called T1. The results for the last type of chains, called T-1, are similar to T1 and are not shown.\\
 The temperature dependence of the T0 chain number and the conductivity are similar. This relation between the conductivity and the T0 chain number seems to indicate that the low temperature metallic behaviour is associated with the T0 chains while the high temperature metallic behaviour is independent of the chain number confirming  the difference  between these two metallic behaviours. This high temperature behaviour probably corresponds to conventional metal, i. e. Fermi liquid. The comparison of T0 and T1 chain numbers for dopings $\delta=0.33$ and $\delta=0.66$ shows that the T1 (and T-1) chains do not contribute significantly to the conductivity. By contrast, the T1 (and T-1) chain number increases with the semiconductor like behaviour as temperature decreases.\\
\subsubsection{Specific heat}
\begin{figure}
 \begin{center}
\includegraphics[width=7cm]{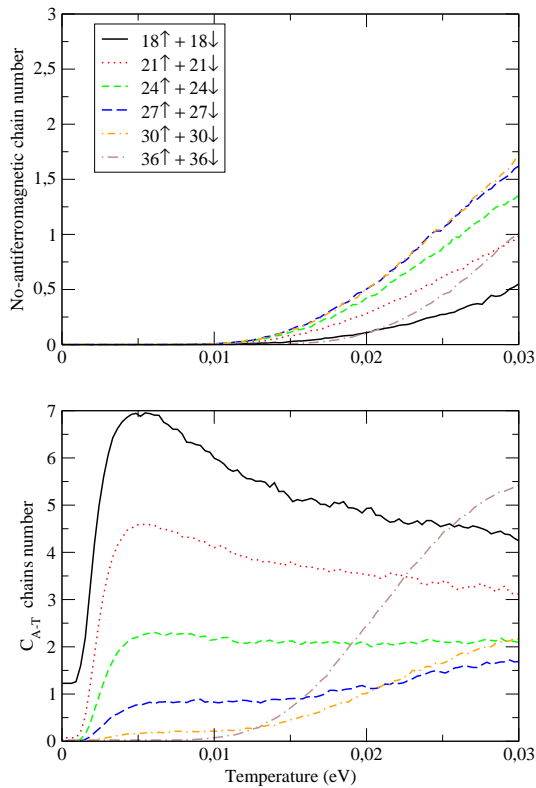}  
 \end{center}
\caption{(Color online). No-antiferromagnetic and $C_{A-T}$ chain numbers versus temperature for different dopings.}
 \label{fig:Antiferro-Triplet}
\end{figure}

 For $\delta=0.16$, $\delta=0.33$ and $\delta=0.5$ the specific heat peaks are associated with the T0 chain formation unlike the peak for $\delta=0$ which is correlated to loop formation. However, all the peaks appear at the same temperature as if they originate from the same phenomenon.\\
 Fig. \ref{fig:Antiferro-Triplet} shows the number of chains which are not antiferromagnetic and the number of antiferromagnetic chains which do not have two copper end sites, i. e. which are not T0, T1 or T-1 chains, these chains are called $C_{A-T}$ chains. The peaks of the $C_{A-T}$ chain number derivatives (not drawn) are centered at the same temperature that the specific heat peaks with a similar shape. We can deduce that these specific heat peaks correspond, partly, to transformation of $C_{A-T}$ antiferromagnetic chains into antiferromagnetic loops for $\delta=0$, and T0 antiferromagnetic chains for the intermediate dopings, during cooling. The $C_{A-T}$ chains are mainly composed by antiferromagnetic chains with one end site being copper site. Indeed the ratio of the number of the chains without copper site end and the number of chains with one copper site end is smaller than $10^{-1}$. Thus, we can consider that the specific heat peaks correspond to ``capture'' of one hole on the second copper site end of each antiferromagnetic chain with one copper site end already occupied by a hole of opposite spin. The resulting T0 chains have an even number of holes.\\
The specific heat peaks can be compared to the oxide superconductor experimental data  \cite{Loram1},\cite{Loram2} which show that the height of the specific heat peaks decrease with the doping when $\delta$ varies from $0.03$ to $0.84$. Our computational results are in good qualitative agreement with these experimental data.\\
\subsubsection{Superconductivity}
Given the sudden conductivity increasing, the superconducting transition is questionable and the role of the T0 chains remains an important question. The maximum low temperature conductivity is not upper than the maximum high temperature conductivity ($kT\simeq0.075\:eV$). The ratio of the maximum of conductivity at low temperature and the minimum is about three. However, the sharp conductivity increasing can point out a tendency to superconductivity which is related to the T0 chains.\\
The coherence of all the results supports the fact that the chosen approximation of conductivity $\Lambda_{xx}$ is relevant.
\section{Conclusion}
\label{sec:conclusion}
This study confirms that the Emery model exhibits a complicated behaviour at low temperature and probably captures  the essential physics of some materials with CuO$_{2}$ plane. However, no sure evidence of superconductivity is found for the interaction parameters used in our simulations. Our results show metal-insulator transitions for $\delta=0$ and $\delta=1$ dopings. For intermediate dopings two different metallic behaviours are evidenced.  A more detailed study shows that the low temperature metallic behaviour is due to antiferromagnetic loops and chains. The physic of these antiferromagnetic chains (or stripes?) seems dominate at low temperature. The antiferromagnetic loops can be related to the circulating currents.\\
As for all simulations in general, we must point out the weakness of this method. The periodic boundary conditions combined with the small size of the system must introduce additional correlations which can modify the physic of the model. Indeed, in the present case, the maximum distance between two sites is about three elementary cells. The slices number is another problem. In principle, the method necessitates many simulations for increasing values of n with the aim of doing extrapolation. Simulations for only one value of n were realize because of the computation duration. Thus the transition temperatures should not be relevant.\\ 
The choice of the sub-systems for the breakup of the system certainly has  an influence. The breakup was chosen in such a way that the number of common sites of two neighbouring systems is minimum. This is in respect with the aim of the method. Moreover, this breakup is certainly the best in accordance with the physic of the CuO$_{2}$ plane. The matrix size is not too large for the computation.\\
Despite these weaknesses and the approximations, this simulation method is the alone method which allows to observe structures like loops and chains in the real two-dimensional array. It gives resuls which seem to relate the circulating current and the stripe concepts. Moreover the behaviour of the specific heat is in good qualitative agreement with the experimental data. Thus the presented simulation results can, perhaps,  suggest a way to supplement the theories proposed to explain the hight $T_{c}$ superconductivity. In this possible scenario the superconductivity is related to the T0 chains, this means that the two holes of the copper end sites of each T0 chain form a Cooper pair. The coherence length is the mean length between the two end sites of the T0 chains. This mean length is about $2.7$ times the lattice parameter (curve not shown). This is in good agreement with the experimental data. The pairing process is the consequence of the particular structure of the  CuO$_{2}$  plane and the repulsive interactions. Indeed, we observe (simulation results not shown) that the low temperature behaviour is strongly infuenced by the values of the interaction parameters $U_{d}$, $U_{p}$ and $V_{dp}$. For example, the low temperature metallic behaviour disappears when $V_{dp}$ is larger than $1.115 eV$, the other parameters being unchanged. This behaviour change is very sudden.\\

\section*{Acknowledgments}
\label{sec:acknow}
We would like to acknowledge useful discussions with V. Ta Phuoc.



\end{document}